\documentclass[10pt,letterpaper,onecolumn]{article}
\usepackage{amsmath,amssymb,graphicx,cite,url,float,dsfont}

\begin{document}
\title{\textsf{\textbf{Single-pixel imaging with origami pattern construction}}}

\author{Wen-Kai Yu$^{1,*}$, Yi-Ming Liu$^2$}

\footnotetext[1]{Center for Quantum Technology Research, School of Physics, Beijing Institute of Technology, Beijing 100081, China. Correspondence and requests for materials should be addressed to W.-K.Y. (email: yuwenkai@bit.edu.cn)}
\footnotetext[2]{Department of Physics, Beijing Normal University, Beijing 100875, China}

\date{}

\maketitle

\renewenvironment{abstract}{%
    \setlength{\parindent}{0in}%
    \setlength{\parskip}{0in}%
    \bfseries%
    }{\par\vspace{12pt}}

\begin{abstract}
Single-pixel compressive imaging can recover images from a small amount of measurements, offering many benefits especially for the scenes where the array detection is unavailable. However, the widely used random patterns fail to explore internal relations between the patterns and the image reconstruction. Here we propose a single-pixel imaging method based on origami pattern construction with a better imaging quality, but with less uncertainty of the pattern sequence. It can decrease the sampling ratio even to 0.5\%, really realizing super sub-Nyquist sampling. The experimental realization of this approach is a big step forward toward the real-time compressive video applications.
\end{abstract}

\noindent Imaging is extremely important for acquiring the light field information of the target \cite{YuOC2017s}. Single-pixel imaging \cite{Edgar2019s,BSun2013s,Radwell2014s,Bian2016s}, as one of popular imaging technologies, can obtain images by using only a point detector without spatial resolution, thus it has extraordinary application prospects, especially at non-visible wavelengths. Except for point-by-point scanning, single-pixel imaging has also developed some imaging techniques based on Fourier \cite{Zhang2015s}or Hadamard transforms \cite{Huynh2016s,MJSunNC2016s,Phillips2017s} (i.e., complete orthogonal basis modulation, full sampling) and correlated/ghost imaging (GI) \cite{Shih1995s,Shapiro2008s,YuOE2014s,Gong2016s,Liu2017s} (i.e., random modulation, oversampling). In recent ten years, with the rapid development of information theory, single-pixel imaging \cite{Baraniuk2008s} based on compressed sensing \cite{Donoho2006s,Candes2006s} has come into being. This approach obtains images of sparse/compressive targets at very low sampling rates, breaking the Nyquist-Shannon sampling limit and bringing dawn of subsampling. However, compressive imaging sacrifices the measurement time and the computational complexity for acquiring spatial resolution, and generally uses random modulation. As a consequence, its sampling rate is still higher than 30\% by a large probability, which has become the bottleneck of compressive imaging for realizing real-time practical applications, especially in large pixel-scale imaging. Recent studies have demonstrated that not all measurements contribute to the reconstructions of computational imaging, resulting in a technology named correspondence imaging (CI) \cite{Luo2011s,Luo2012s,Shih2011s,YuCPB2015s,MJSunAO2015s}, where only the measurements well above or below the mean of detected values are used for reconstruction. Now that there exists such a phenomenon, we are thinking, why not just construct a universal deterministic (instead of random) measurement matrix where the modulated patterns with the most contributions to the reconstruction are preferentially displayed. In particular, an earlier research work \cite{MJSunSR2017s} reported a ghost imaging method based on ``Russian Dolls" (RD) ordering of the Hadamard basis, which provided us with inspiration.

In this Letter, we propose an origami model to construct deterministic patterns, decreasing the sampling ratio to 0.5\%, which is $\sim2$ orders of magnitude lower than that of traditional compressive imaging. The row vectors stretched from each constructed pattern are orthogonal to each other, suitable for large-pixel-scale reconstruction. Therefore, it opens a door to the super subsampling applications for compressive video \cite{Zheng2009s,Yu2018s}.

Now let us begin with the definition of differential ghost imaging (DGI) \cite{Gatti2010s}, one method that can greatly improve the quality of conventional GI. The bucket/single-pixel signal can be defined as $S_B=\iint I_B(x_o,y_o)T(x_o,y_o)dx_ody_o$, where $I_B(x_o,y_o)$ and $T(x_o,y_o)$ represents the intensity and transmission function at the spatial position $(x_o,y_o)$ of the object $x$ of $p\times q=N$ pixels, respectively. Similarly, the reference bucket signal can also be defined as $S_R=\iint I_R(x,y)dxdy$, where $I_R(x,y)$ stands for the values at the spatial position $(x,y)$ of the reference speckle patterns. By using a differential bucket signal $S_{\Delta}=S_B-\frac{\langle S_B\rangle}{\langle S_R\rangle}S_R$ instead of $S_B$ in second-order correlation $\langle S_B I_R(x,y)\rangle$, where $\langle\cdots\rangle$ signifies the ensemble average, we can compute a differential ghost image $\widetilde{U}_\textrm{DGI}$ from $M\gg N$ measurements via $\widetilde{U}_\textrm{DGI}=\langle S_{\Delta}I_R(x,y)\rangle=\langle S_B I_R(x,y)\rangle-\frac{\langle S_B\rangle}{\langle S_R\rangle}\langle S_R I_R(x,y)\rangle$. A high image quality can be guaranteed by DGI, but at the cost of oversampling.

Later, another method called CI \cite{Luo2011s,Luo2012s,Shih2011s,YuCPB2015s,MJSunAO2015s} was reported, in which a positive (or negative) ghost image can be retrieved by only averaging some small fractions of the reference subset $I_R^+$ (or $I_R^-$), i.e., $\left\langle{{I_{R_+}}}\right\rangle$ (or $\left\langle{{I_{R_-}}}\right\rangle$), corresponding to $\{S_B^+|S_B\gg\left\langle S_B\right\rangle\}$ (or $\{S_B^-|S_B\ll\left\langle S_B\right\rangle\}$). It can be seen that CI just need to calculate addition instead of second-order correlation, so CI takes less computation time than DGI. Although only a small number of patterns are involved in the calculation, the number of measurements is not decreased, and the conditional averaging of the reference patterns is made after oversampling. But it still can be concluded that not all the measured bucket/single-pixel intensities are necessary for reconstruction. Since single-pixel compressive imaging shares the same mathematical model with ghost imaging, this conclusion can be also applicable to compressive imaging. Obviously, the random patterns are not the best choices to minimize the number of measurements, because such a pattern has a high probability of transmitting or reflecting light at half of all pixels. It is really difficult for us to determine which pattern has a higher reconstruction contribution or produces a higher bucket intensity. However, we find that, if we encode on the spatial light modulator (SLM) with a pattern consisting of all-one, which can be treated as a connected domain, the detected intensity value will be the highest. If the pattern is divided equally into two connected domains with the white-black pixels 50\%-50\%, then the detected value will be the second highest with a high probability. On this basis, the problem can be partially deduced to the contributions of the piece number of connected domains (CDs) of patterns to the image quality. Here we propose a novel origami pattern construction method, which makes full use of symmetric reverse folding (reverse the values on the corresponding pixels at the symmetric positions) and the axial symmetry of the rescaled pattern. The steps of our origami pattern construction are as follows.

\textbf{Step 1:} Assume that the pixel-sizes of patterns are all $p\times q=n$ square matrices, with the same size as the object, where $p$, $q$ ($p=q$) are all even numbers. Suppose there are $n$ patterns in total, we divide them into $n/4$ groups, each of which with four patterns. The group number is denoted by $i=1,2,3,\cdots,n/4$. Note, as an initial setting, the first pattern $P_{1,1}$ in the first group is a matrix with all ones. Then the second and the third patterns in this group $P_{1,2}$, $P_{1,3}$ are obtained by inversely folding the first one $P_{1,1}$ in this group up and down, left and right, respectively. That is, remain the upper (or the left) half unchanged while the lower (or right) half is axisymmetric with the upper (or left) half but with its values in the lower (or right) half taking the opposite. By this means, we realize black (-1) and white (+1) inverting. The fourth pattern $P_{1,4}$ in this group is formed by performing both up-down and left-right reverse folding operations (the sequence of operations is interchangeable) on the first pattern $P_{1,1}$. Similarly, for all the subsequent groups, the rest three patterns in the $i$th group, $P_{i,2}$, $P_{i,3}$ and $P_{i,4}$ will be generated from $P_{i,1}$ following the same processes.

\begin{figure}[htbp]
\centering
\includegraphics[width=0.95\linewidth]{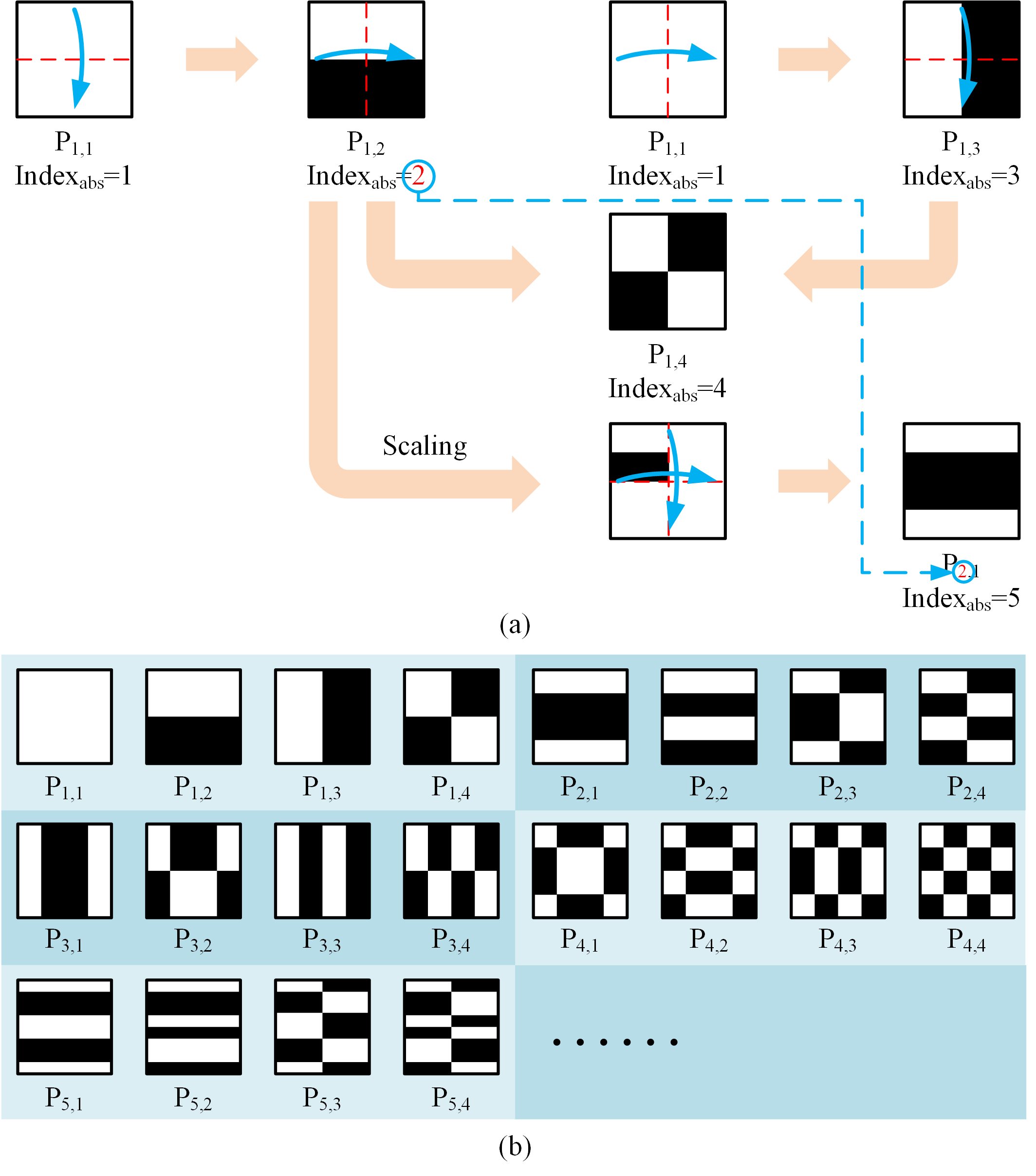}
\caption{Origami pattern construction. (a) Pattern forming process for Steps 1-2. (b) Results obtained after Step 2.}
\label{fig:origamisteps12}
\end{figure}

\textbf{Step 2:} The first pattern $P_{i,1}$ in the $i$th group (also the $4(i-1)+1$th pattern in the complete sequence) is built on the basis of the $i$th pattern in the current complete sequence. Take $P_{2,1}$ in the second group for example, we firstly scale both horizontal and vertical pixel dimensions of the second pattern in the complete sequence (i.e., $P_{1,2}$) to their $\frac{1}{2}$ size, and place the scaled one on the upper left $1/4$ part of a $p\times q$ square matrix with all zeros. Secondly, we perform the up and down, left and right axial symmetry about the midline lines on the vertical and horizontal axes of the matrix. After that, the pattern $P_{2,1}$ (also the 5th pattern in complete sequence) is generated. Then, repeat Step 1 to acquire the second to fourth patterns in the second group, $P_{2,2}$, $P_{2,3}$ and $P_{2,4}$ (also the 6th to 8th patterns in complete sequence). By analogy, we can create all $n/4$ groups. Fig.~\ref{fig:origamisteps12}(a) gives a good illustration of Steps 1 and 2, and Fig.~\ref{fig:origamisteps12}(b) shows the result obtained after Step 2.

\begin{figure}[htbp]
\centering
\includegraphics[width=0.95\linewidth]{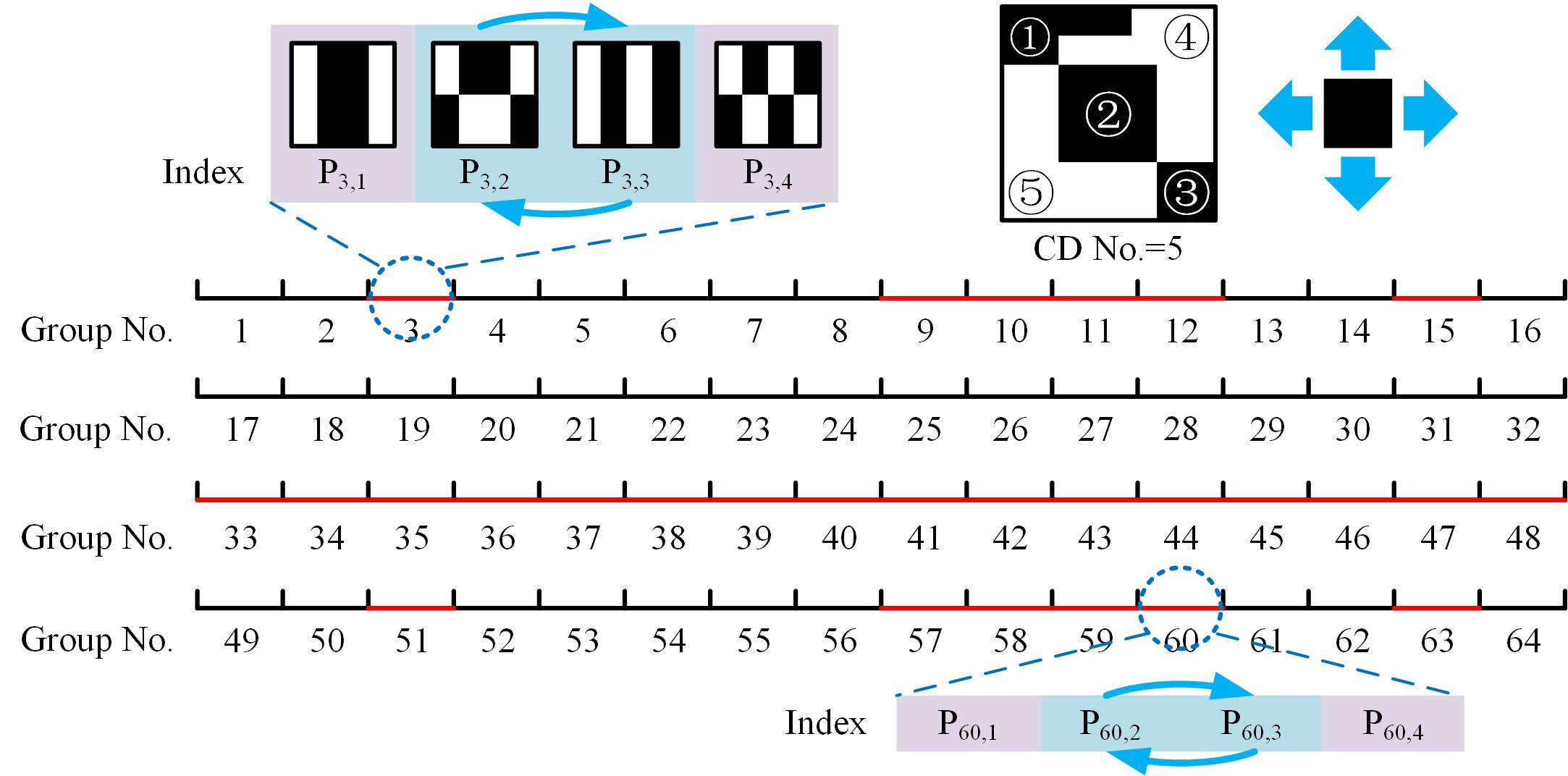}
\caption{Group numbers for cases of $n=64,\ 256$. The patterns $P_{i2}$ and $P_{i3}$ with red marked group numbers need to be order exchanged. The subgraph in the upper right corner illustrates an example of the connected domains with a number of 5.}
\label{fig:CDalternate}
\end{figure}

\textbf{Step 3:} Adjust the pattern sequence to ensure that the number of CDs for each pattern in the $i$th group is in an incremental order. Here a CD is defined as an up-down-left-right connected area consisting of equal values (see the upper right corner of Fig.~\ref{fig:CDalternate}). The neighborhoods on both sides of the axis may cause the partition of CDs, which is worth of investigating. Let's first look at the pattern $P_{i,1}$ whose pixel values on both sides of the symmetry axis are the same, its contribution to the recovery is the largest in the $i$th group. Since the pattern $P_{i,4}$ needs double symmetric reverse folding, its number of CDs is largest (with lowest contribution) in the $i$th group. The patterns $P_{i,2}$ and $P_{i,3}$ are produced by only once symmetric reverse folding, whether to first perform up-down or left-right operation is approximately equivalent, thus they are particularly worth considering. As shown in Fig.~\ref{fig:CDalternate}, the patterns $P_{i,2}$ and $P_{i,3}$ with a group indication (ID) (note: not the absolute subscript) $i=3,\ 9\sim12,\ 15$ need to be order reversed for the case of $n=64$, while those with a group ID $i=3,\ 9\sim12,\ 15,\ 33\sim48,\ 51,\ 57\sim60,\ 63$ require adjustments for the case of $n=256$. This rule also exists in higher-order cases. For this, we set fourfold as a level. From top to bottom, the total $n/4$ groups (as a whole) can be catalogued into four parts: only the second part stays still, the groups in the third part all need to be adjusted, while the groups in the first and fourth parts also need to be adjusted but with the same group IDs, depending on the recursive layer. Then repeat the operation for the first part (as a new whole) until the number of groups in each new quarter part equals to 1. In the last layer, only the third part needs to be adjusted. Then, switch the order of the patterns $P_{i,2}$ and $P_{i,3}$ in these found groups.

\begin{figure}[htbp]
\centering
\includegraphics[width=0.95\linewidth]{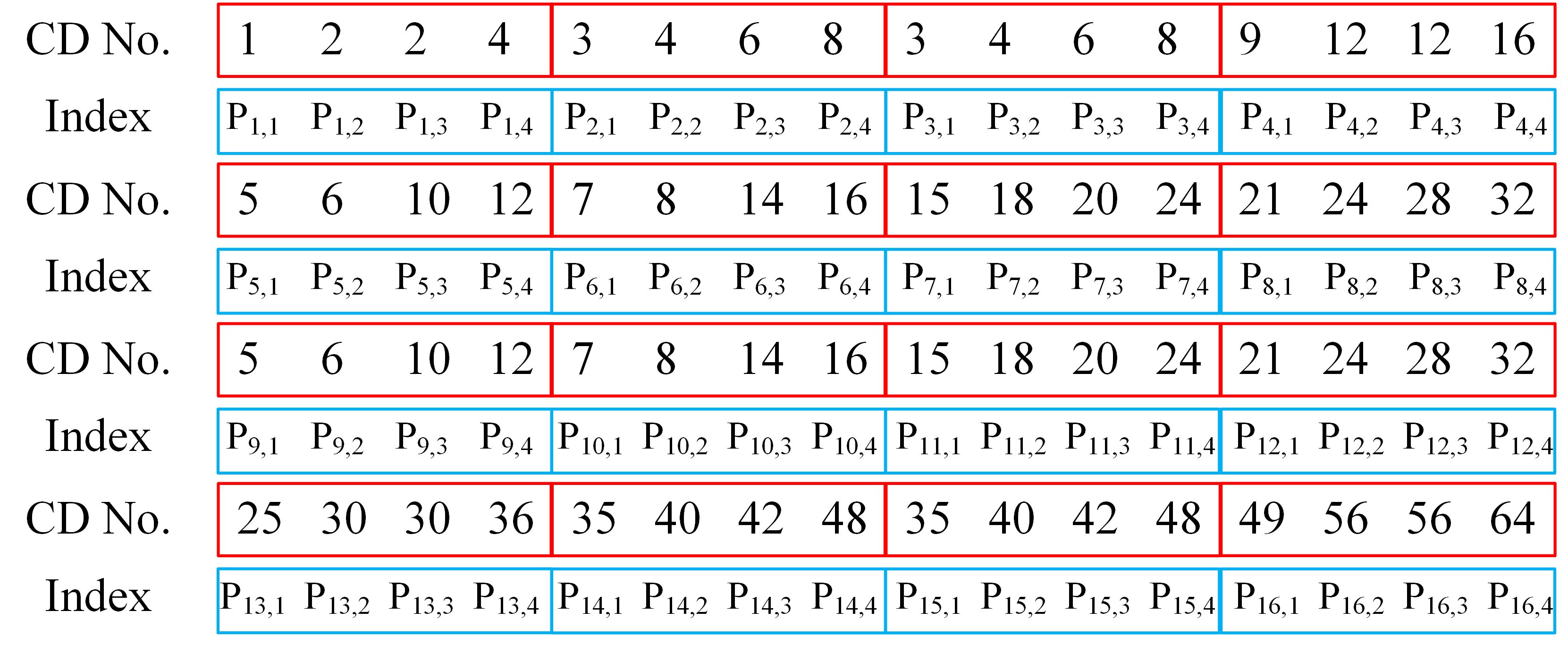}
\caption{Numbers of CDs for a final pattern sequence of $n=64$.}
\label{fig:finalpatternsnum}
\end{figure}

\begin{figure}[htbp]
\centering
\includegraphics[width=0.95\linewidth]{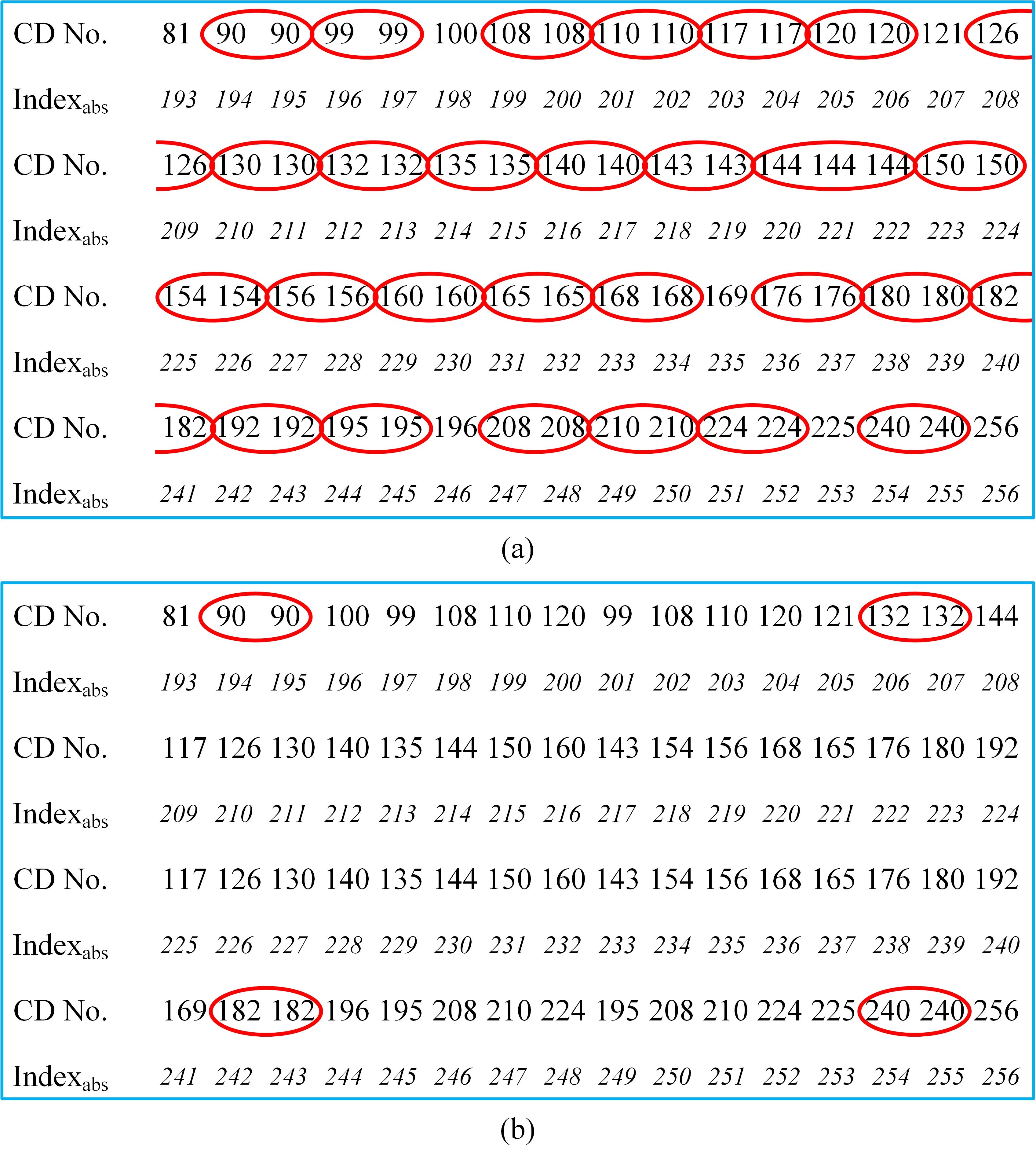}
\caption{Comparison of the number of CDs between Russian dolls (RD) ordering of the Hadamard basis (a) and our method (b) in high-order part. The patterns with the same CDs are circled by red ellipses.}
\label{fig:comparison}
\end{figure}

If we perform left-right and up-down reverse folding to generate $P_{i,2}$ and $P_{i,3}$ in Step 1, respectively, then the ID positions should be adjusted accordingly in Step 3. After the above three steps, we will get the final pattern sequence. Fig.~\ref{fig:finalpatternsnum} gives an example of the final pattern sequence for $n=64$. In order to deeply analyze the advantages of our method, we compare it with the RD method. It is interesting to find that these two approaches have similar numbers of CDs in low-order part (especially for the first 16 patterns, they are the same). Since their subsequent results are both derived from these first 16 patterns, it is advisable to set each 16 patterns as a comparison unit. Fig.~\ref{fig:comparison} shows the differences in their high-order parts (the last 64 patterns in 256-order sequence which also comprises the largest division of the RD method). As shown in Fig.~\ref{fig:comparison}(a), the RD ordering becomes very rough in this part, and has a lot of pattern pairs with the same number of CDs in each comparison unit. Because the maximum partition length of RD is $n/4$ while its minimum segmentation length is 4, which will inevitably incur too much uncertainty. Although the internal order exchange in these pattern pairs does not affect the RD sorting too much, it does bring hundreds of millions possibilities of finding the optimal pattern sequence to get the best imaging quality. While in our scheme, each four patterns are treated as a group, i.e., the partition length is always 4. As shown in Fig.~\ref{fig:comparison}(b), there are only four red marks in our method for the high-order part, with only $2^4$ uncertainty. It makes the reconstructed images more accurate. Thus our method can be regarded as a better strategy.
\begin{figure}[htbp]
\centering
\includegraphics[width=0.95\linewidth]{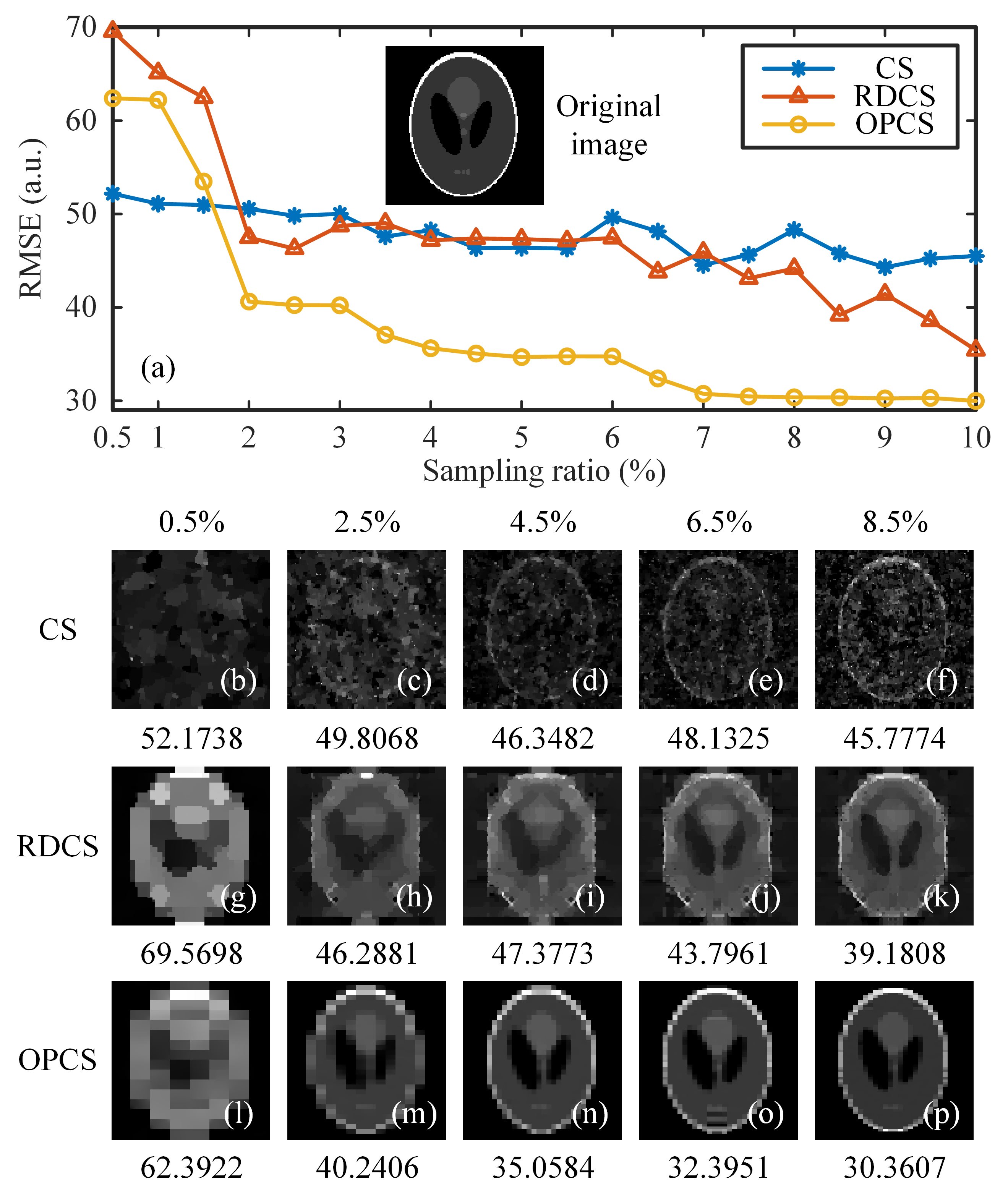}
\caption{RMSE comparisons between CS, RDCS and OPCS. (a) gives an original head phantom image of $128\times128$ pixels and multiple RMSE curves of the sampling ratio. (b)--(f), (g)--(k) and (l)--(p) are CS, RDCS and OPCS reconstructions of $128\times128$ pixels under 0.5\%--8.5\% sampling ratios, respectively. The digits below (b)--(p) are RMSE values.}
\label{fig:simulation}
\end{figure}

Now, each pattern $I_R$ can be sequentially reshaped into a row vector of $1\times n$, and then $n$ such row vectors constitute a full-rank square matrix, which happens to be an orthogonal matrix. It is quite suitable for forming the measurement matrix $A\in\mathds{R}^{m\times n}$ of CS by just selecting the first $m$ row vectors. Here we use a TVAL3 solver \cite{CBLi2010s} to recover the image. For a fair comparison, we combine the RD ordering of the Hadamard basis with CS, instead of second-order correlation used in the original version \cite{MJSunSR2017s}. Here we introduce an unitless performance quantitative measure, the root mean square error (RMSE), which is defined as $\textrm{RMSE}=\sqrt{\frac{1}{pq}\sum\nolimits_{i,j=1}^{p,q}[\tilde{U}(i,j)-U_o(i,j)]^2}$. It describes the square root of differences between the recovered image $\tilde U$ and the original image $U_o$ for all pixels. Generally, the smaller the RMSE value, the better the quality of the image recovered. Since it is a pixel-by-pixel calculation, it cannot provide a very accurate characterization of performance. Therefore, in addition to the RMSE curves of CS, Russian dolls compressed sensing (RDCS) and origami pattern compressed sensing (OPCS), we also present their recovered results under different sampling ratios from 0.5\%--8.5\% with a 2\% stepping increase, as shown in Fig.~\ref{fig:simulation}. It is obvious that the performance of our method outperforms those of CS and RDCS. Only 2.5\% measurements are enough for OPCS to generate a nice reconstruction, and its limit sampling rate can reach 0.5\%.

\begin{figure}[tbp]
\centering
\includegraphics[width=\linewidth]{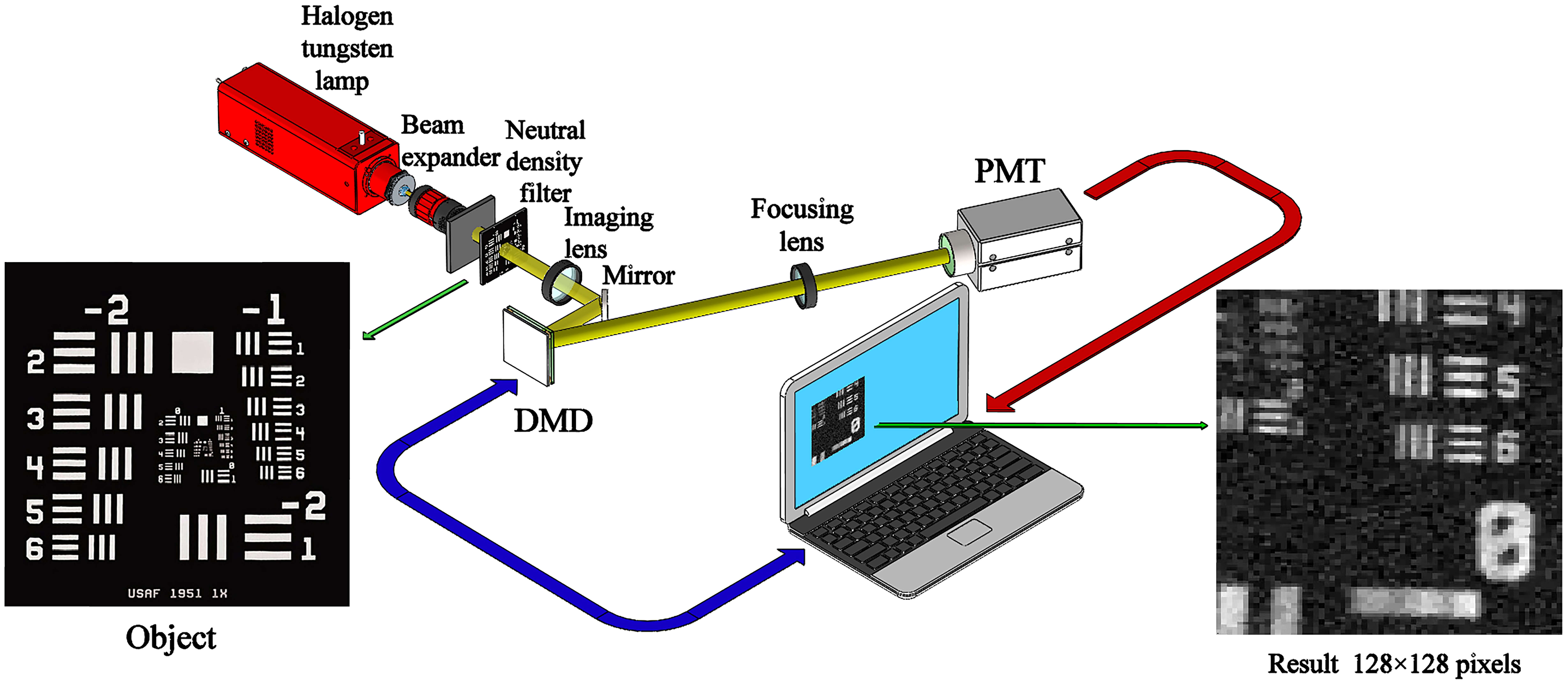}
\caption{Schematic of the experimental setup. After being collimated and broadened, the light illuminates the object (a negative 1951 USAF resolution test chart of 3 inch$\times$3 inch) and then it is imaged on the DMD. A PMT collects the differential photon counts via a focusing lens. The reconstructed image of $128\times128$ pixel-size is given in the lower right corner.}
\label{fig:expandresults}
\end{figure}

Our experimental realization is based on a conventional single-pixel camera, as shown in Fig.~\ref{fig:expandresults}. The thermal light from a stabilized halogen tungsten lamp with a wavelength of 360$\sim$2600~nm is collimated and attenuated via a beam expander and a neutral density filter (NDF), respectively. Here the NDF is used to attenuate the light to the ultra-weak light level. After that, the light passes through a negative 1951 USAF resolution test chart, and then it is imaged onto a digital micromirror device (DMD) via an imaging lens and a mirror. The applied DMD consists of $1024\times768$ micromirrors, each of which can be switched between two directions of $\pm12^\circ$, corresponding to 1 and 0. Since the values of our patterns are either 1 or $-1$, we need to divide each generated pattern $I_R$ into a complementary matrix pair \cite{YuSR2014s,YuAO2015s,YuOC2016s} $\hat{I}_R=(I_R+1)/2$ and $\check{I}_R=1_{N\times N}-\hat{I}_R$, where $I_R=\hat{I}_R-\check{I}_R$, and $1_{N\times N}$ stands for an all 1 matrix. Then, these complementary matrices are sequentially encoded on the DMD. A counter-type Hamamatsu H10682-210 PMT is placed on one of the reflection orientations to accordingly make a differential measurement of two adjacent total photon counts. By using OPCS algorithm, we can acquire a very satisfactory image quality, as shown in the illustration of Fig.~\ref{fig:expandresults}.

In conclusion, a single-pixel compressive imaging method with origami pattern construction is proposed and demonstrated experimentally with differential modulation. In our scheme, following the fluctuation characteristic of measurement values, the patterns with the most contributions to the recovery should be preferentially displayed. By symmetric reverse folding, axial symmetry and partial pattern order adjustment, the generated deterministic patterns ensure a better image quality but with less uncertainty of the pattern sequence, compared with traditional CS and RDCS. The sampling ratio can be reduced to 0.5\%, even about 2 orders of magnitude lower than that of traditional CS, actually realizing super sub-Nyquist sampling. Therefore, our scheme can be used to greatly improve real-time performance of single-pixel compressive video applications, especially where array detectors are unavailable.

\section*{\textsf{Funding}}
This work is supported by the National Natural Science Foundation of China (61801022), the Beijing Natural Science Foundation (4184098), the National Key Research and Development Program of China (2016YFE0131500), the International Science and Technology Cooperation Special Project of Beijing Institute of Technology (GZ2018185101), and the Beijing Excellent Talents Cultivation Project - Youth Backbone Individual Project.

\section*{\textsf{Acknowledgment}}
The authors warmly appreciate Shuo-Fei Wang and Ya-Xin Li for preparing Fig.~\ref{fig:expandresults}.

\bigskip





\end{document}